\documentclass[useAMS, usenatbib]{mn2e}

\usepackage{times}
\usepackage{graphicx, bm, amssymb, xcolor}
\usepackage{verbatim}
\usepackage[backref,breaklinks,colorlinks,citecolor=blue]{hyperref}
\usepackage[all]{hypcap}

\newcommand{\noopsort}[1]{}

\newcommand{\msun}{M_\odot}

\definecolor{darkgreen}{rgb}{0.13, 0.55, 0.13}

\begin{document}

\title[Core Structure in Radiation-MHD]{What Physics Determines the Peak of the IMF? Insights from the Structure of Cores in Radiation-Magnetohydrodynamic Simulations}

\author[Krumholz et al.]{Mark R.~Krumholz$^1$\thanks{mark.krumholz@anu.edu.au}, Andrew T.~Myers$^2$, Richard I.~Klein$^{3,4}$, and
Christopher F.~McKee$^{4,5}$
\\ \\
$^1$ Research School of Astronomy \& Astrophysics, Australian National University, Canberra, ACT, Australia\\
$^2$ Lawrence Berkeley National Laboratory, Berkeley, CA USA\\
$^3$ Lawrence Livermore National Laboratory, Livermore, CA, USA\\
$^4$ Department of Astronomy, University of California, Berkeley, Berkeley, CA, USA\\
$^5$ Department of Physics, University of California, Berkeley, Berkeley, CA, USA\\
}

\date{\today}

\pagerange{\pageref{firstpage}--\pageref{lastpage}} \pubyear{2015}

\maketitle

\label{firstpage}

\begin{abstract} 
As star-forming clouds collapse, the gas within them fragments to ever-smaller masses. Naively one might expect this process to continue down to the smallest mass that is able to radiate away its binding energy on a dynamical timescale, the opacity limit for fragmentation, at $\sim 0.01$ $M_\odot$. However, the observed peak of the initial mass function (IMF) lies a factor of $20-30$ higher in mass, suggesting that some other mechanism halts fragmentation before the opacity limit is reached. In this paper we analyse radiation-magnetohydrodynamic simulations of star cluster formation in typical Milky Way environments in order to determine what physical process limits fragmentation in them. We examine the regions in the vicinity of stars that form in the simulations to determine the amounts of mass that are prevented from fragmenting by thermal and magnetic pressure. We show that, on small scales, thermal pressure enhanced by stellar radiation heating is the dominant mechanism limiting the ability of the gas to further fragment. In the brown dwarf mass regime, $\sim 0.01$ $M_\odot$, the typical object that forms in the simulations is surrounded by gas whose mass is several times its own that is unable to escape or fragment, and instead is likely to accrete. This mechanism explains why $\sim 0.01$ $M_\odot$ objects are rare: unless an outside agent intervenes (e.g., a shock strips away the gas around them), they will grow by accreting the warmed gas around them. In contrast, by the time stars grow to masses of $\sim 0.2$ $M_\odot$, the mass of heated gas is only tens of percent of the central star mass, too small to alter its final mass by a large factor. This naturally explains why the IMF peak is at $\sim 0.2$ $M_\odot$.
\end{abstract}

\section{Introduction}
\label{sec:intro}

The origin of the stellar initial mass function (IMF) is one of the outstanding problems in contemporary theoretical astrophysics \citep{krumholz14c, offner14a}. The observed IMF displays a characteristic peak at $\sim 0.2-0.3$ $M_\odot$, with a decline in the number of objects on either side of this plateau \citep{kroupa02c, chabrier03a, chabrier05a, bastian10a, parravano11a, offner14a}. This peak appears to be universal or nearly so within the Milky Way, and even in the most extreme environments to which we have access its location is different by at most a factor of $\sim 2-3$ \citep[e.g.,][]{van-dokkum10a, van-dokkum11a, cappellari12a, spiniello12a, spiniello15a, conroy13a}.

The origin of this mass scale is far from clear. Stars form in a turbulent, magnetised, radiating medium. Given this complexity, it is not surprising that a number of theoretical models have been proposed, emphasising different physical mechanisms as providing the key element. For example, some authors propose that the location of the peak is set by the thermal Jeans mass in nearly-isothermal star-forming clouds, in which case the key physical process is whatever determines the mean density and temperature of the isothermal gas \citep[e.g.,][]{larson92a, bate05a, clark05a, bonnell06a}, and neither turbulence, nor magnetic fields, nor radiative transfer are important processes. 

Others add turbulence to this picture, proposing that the peak of the IMF is instead set by the Jeans mass evaluated at a characteristic density set by turbulent compression \citep[e.g.,][]{hennebelle08b, hennebelle09a, hennebelle13a, hopkins12d, hopkins13a, hopkins13e}. Some authors suggest that magnetic fields and the support they provide have an important role in limiting how gas fragments, either because magnetic pressure changes shock jump conditions by limiting compression \citep[e.g.,][]{padoan02a, padoan07a, gong11b} or because magnetic pressure directly suppresses small-scale fragmentation \citep{hennebelle08a, hennebelle11a}. Yet other authors have emphasised the role of radiative processes in suppressing fragmentation below a critical mass, thereby picking out the location of the IMF peak that way (e.g., \citealt{whitworth98a, larson05a, jappsen05a, bonnell06a}).

In particular, \citet{jappsen05a} and \citet{bonnell06a} both performed experiments with parameterised equations of state intended to mimic the effects of radiative processing, and showed that the location of the IMF peak scales directly with the Jeans mass evaluated at the density where their equations of state stiffen, strongly suggesting a role for radiative processes in determining the characteristic stellar mass. However, in these simulations the gas temperature and thus the effective equation of state was not calculated self-consistently, and the calculations did not include the effects of radiation from the stars themselves. That this radiation is in fact the dominant mechanism in determining the gas temperature structure was first pointed out analytically by \citet{krumholz06b} and numerically by \citet{bate09a}, and both analytic models and simulations locating the origin of the IMF peak in stellar radiative feedback have been published by a number of authors \citep[e.g.,][]{offner09a, bate09a, bate12a, bate14a, krumholz11e, krumholz11c, krumholz12b, guszejnov16a}.

On top of all these processes, the characteristic mass scale can be shifted lower by both protostellar outflows \citep{hansen12a, krumholz12b} and strong magnetic fields. The former eject mass directly, while the latter lower accretion rates and inhibit the formation of massive collapsing regions \citep{li10b, hocuk12a}.\footnote{Once a collapsing region does form, however, the field appears to have the opposite effect, suppressing fragmentation and favouring formation of more massive stars \citep{hennebelle11a, commercon11c, myers13a}.} 

The difficulty of teasing out the physics that is responsible for setting the IMF is partly driven by the fact that most simulations to date do not include all the possibly-important effects. While there are a large number of simulations including turbulence and gravity, there are relatively few that also include magnetic fields, radiative transfer, and protostellar outflows, and even fewer that combine all of these elements. The first published radiation-hydrodynamic simulations of star cluster formation (as opposed to formation of individual stars or small multiple systems) by \citet{bate09a, bate12a}, \citet{offner09a}, and \citet{krumholz11c}, included neither magnetic fields nor protostellar outflows. \citet{hansen12a} and \citet{krumholz12b} included outflows with radiation, but not magnetic fields. The only published studies reporting radiation-magnetohydrodynamic simulations of the formation of multiple stars are those of \citet{price09a}, \citet{commercon11a}, \citet{peters11a}, and \citet{myers13a, myers14a}. Only the last of these both includes protostellar outflows and, most importantly, forms enough stars that one can obtain meaningful statistics from it, albeit only in the mass range near the IMF peak that is best sampled. Thus these simulations are unique in their ability to compare the relative importance of magnetic fields and radiation in determining how gas fragments, and to do so including the effects of outflows, which tend to weaken the influence of radiative feedback \citep{hansen12a, krumholz12b}.

In this paper we use the simulations of \citet{myers14a} to study what physical processes are responsible for determining how the gas fragments, and thus for setting the location of the IMF peak. Our strategy is to examine in detail the gas in the vicinity of each forming star, starting from the instant at which a collapsed object appears and following as it accretes, with the goal of measuring the relative importance of magnetic and thermal support, and, for the latter, the importance of radiative effects in raising or lowering the level of thermal support. This zoom-in on cores approach is quite similar to that employed by \citet{bonnell04a} and \citet{smith09a}, with the difference that we have access to simulations that include a much wider range of physical processes, enabling much stronger conclusions.

The remainder of this paper is organised as follows. In \autoref{sec:analysis} we review the properties of the simulations and then discuss our analysis method. We present the results of our analysis in \autoref{sec:results}, and discuss their implications in \autoref{sec:discussion}. We summarise and conclude in \autoref{sec:conclusion}.

\section{Analysis}
\label{sec:analysis}

\subsection{Summary of the Simulations}

We analyse the simulations published by \citet{myers14a}. We refer readers to that paper for a full description of the simulations, and here simply summarise details that will become relevant below. All simulations include radiative transfer (including stellar radiation feedback), and begin from initial conditions produced by driving turbulence in a periodic box in order to let it reach statistical equilibirum before turning on self-gravity. \citeauthor{myers14a}~considered three different initial magnetic field strengths: no magnetic field (referred to as the ``hydro" case), a ``weak" field case with an initial mass to flux ratio equal to 10 times the critical value for gravitational collapse, and a ``strong" field case with an initial mass to flux ratio set to twice the critical value for collapse. We use the high-resolution versions of these simulations, which have finest cells of 23 AU. Once gas is Jeans unstable even at this resolution, we replace it with an accreting sink particle \citep{krumholz04a} that is coupled to a protostellar evolution calculation \citep{offner09a} and injects radiation and winds \citep{cunningham11a} back into the computational domain. The accretion process removes mass but not magnetic flux from the computational domain, and thereby decreases the mass to flux ratio inside the accretion region around the sink particle. Sink particles track the angular momentum of the material they accrete, and thus have a well-defined spin axis. All simulations use a computational domain containing 1000 $\msun$ of gas in a periodic domain $0.46$ pc on a side, giving an initial mean density of $n_{\rm H} = 3.0\times 10^5$ H nuclei cm$^{-3}$, and a column density $\Sigma = 4700$ $M_\odot$ pc$^{-2} = 1$ g cm$^{-2}$. The initial gas temperature is 10 K, and the initial magnetic field strengths are $0.16$ and $0.81$ mG in the weak and strong field runs, respectively. By the end of these simulations, the hydro, weak, and strong runs have formed 100, 74, and 70 stars, respectively.

We note that, due to the high numerical costs of radiation-magnetohydrodynamics, the simulations of \citet{myers14a} do not run to the end of the star formation process. As a result, new stars were continuing to appear, and those stars that formed at earlier times were continuing active accretion, and had not yet reached their final masses. Consequently, the median mass was somewhat smaller than the observed peak of the IMF. This median mass resulted from the balance between the growth of existing stars to higher masses and the formation of new stars at the bottom of the mass distribution. Once the formation of new stars ceases or tapers off, the median mass will have to rise, unless for some reason accretion onto existing stars stops at the same time; we show in  \autoref{ssec:accretion_stop} that such a rapid halt to accretion is implausible. For a population is still forming, the relevant comparison to observations is not with the IMF but with the \textit{protostellar} mass function \citep{mckee10c, offner11a}, which is the mass function expected for a population of class 0 and class 1 protostars that will end up with a mass distribution that follows the IMF. \citet{myers14a} show that the mass distribution produced in these simulations is in good agreement with the protostellar mass function expected for the observed IMF. Since we are interested in precisely the question of how a protostellar mass function transforms into the IMF, this makes the simulations a particularly useful vehicle for analysis.

\subsection{Core Profiles and Critical Masses}
\label{ssec:prof}

\begin{figure}
\centerline{
\includegraphics[width=0.85\columnwidth]{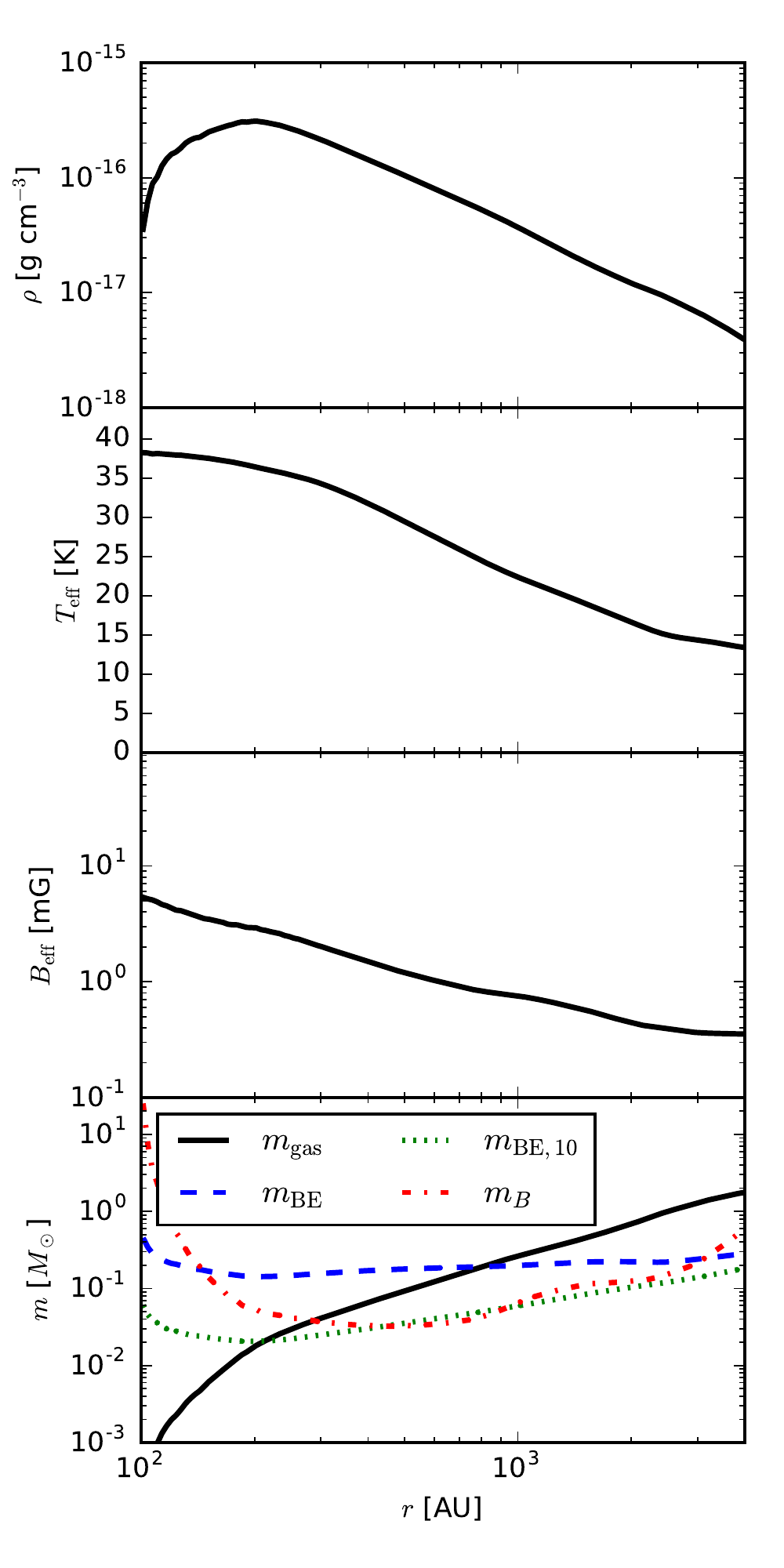}
}
\caption{
\label{fig:coreprof_example}
Example of the density, effective temperature, effective magnetic field strength, and mass profiles around a $0.17$ $M_\odot$ protostar, in a simulation with 23 AU resolution. The top panel show the mean density of the material enclosed within each radius. The next two panels show the temperature and magnetic field strength that would yield values equal to the mass-weighted mean sound speed and mean magnetic flux interior to that radius, respectively. In the final panel, the solid line shows the mass of gas enclosed, the thick dashed line shows the Bonnor-Ebert mass computed from the mass-weighted mean sound speed and mean density, the dotted line shows the Bonnor-Ebert mass computed from the mean density using a fixed gas temperature of 10 K, and the dot-dashed line shows the magnetic critical mass.
}
\end{figure}

We are interested in determining the role of thermal pressure (both with and without radiative transfer effects) and magnetic support in inhibiting fragmentation in the collapsing regions in our simulations. To this end, we identify every sink particle at each output time slice, and use the \texttt{yt} software package \citep{turk11a} to compute a series of quantities. We examine 128 concentric spherical shells centred on each sink particle, with the inner edge of the innermost shell placed at a distance of 100 AU from the particle and the most distant at 4000 AU.\footnote{We consider spherical shells rather than cylinders, and below we consider support against fragmentation in 3D rather than 2D geometry, because for the most part our stars are not surrounded by large disks as a result of magnetic braking. This is consistent with the general finding of MHD simulations that magnetic fields at realistic strengths either, depending on the problem setup, prevents disk formation entirely, or reduces the sizes of disks to tens of AU -- see the recent review by \citet{li14a}. Such disks are too small for us to resolve given our 23 AU resolution.} For each shell, we compute the total gas mass enclosed $m_{\rm gas}$ (excluding the sink particle), the mean density $\rho$, and, in order to assess the amount of thermal support, the mass-weighted mean isothermal sound speed $c_s$. We compute these quantities both cumulatively, meaning that we take the mass and mean sound speed of all gas within the shell, and differentially, meaning that we consider only the material between two shells. We show in \autoref{app:differential} that the results for using either method are qualitatively the same, and so for the remainder of the paper we will focus on the cumulative quantities, which are somewhat less noisy. From $\rho$ and $c_s$, we compute the Bonnor-Ebert mass \citep{ebert55a, bonnor56a},
\begin{equation}
\label{eq:mbe}
m_{\rm BE} = 1.86 \sqrt{\frac{c_s^3}{G^3\rho}},
\end{equation}
and we therefore have $m_{\rm BE}$ as a function of radius around each sink particle. Note that the coefficient here is $1.86$ rather than the more familiar $1.18$ because we are using the mean density rather than the surface density; in the isothermal case, this is a factor of $2.465$ higher \citep{mckee99b}, explaining the increased coefficient. The Bonnor-Ebert mass characterises the level of thermal support in the gas; objects with a mass less that $m_{\rm BE}$ are stable against collapse. Note that, since the Bonnor-Ebert mass has been computed for isothermal gas (or more generally for polytropic gas -- \citealt{mckee99b}), while our gas is neither isothermal nor polytropic, this calculation of the mass that can be supported is only approximate. Nonetheless, it should provide a useful estimate of the importance of thermal pressure support.

To assess the importance of radiative heating by stars, we repeat the computation of the Bonnor-Ebert mass with the sound speed fixed to $c_s = 0.19$ km s$^{-1}$, the sound speed for molecular gas at 10 K, the background temperature in the simulations, and the temperature that the gas would have in the absence of radiative heating. We refer to this quantity as $m_{\rm BE,10}$. Note that this is an imperfect proxy for the effects of radiative heating, because the densities that go into $m_{\rm BE,10}$ have still been derived from a simulation including radiative heating and its effects on the dynamics. Compared to what would be obtained in a purely isothermal solution, this should generally produce lower $m_{\rm BE,10}$, because the increased pressure produced by radiative heating will tend to lower densities. Thus if anything our method underestimates the effects of radiative heating on supporting the gas.

Finally, the key quantity in determining magnetic fields' ability to inhibit fragmentation is the flux $\Phi$. 
We record $\Phi$ over a series of 128 circular areas centred on the sink particle, with the outer edges of the circles lying on the spheres used in the computation of the enclosed mass and $m_{\rm BE}$. We perform this operation on circular areas with 12 different orientations in space, one aligned with the angular momentum vector of the sink particle, and the remaining 11 distributed uniformly following the \texttt{HEALPix} pixelisation scheme \citep{gorski05a}. For each circular area we record the flux in whichever orientation produces the largest value, and we compute from the absolute value of the magnetic field dotted with the surface normal, $|\mathbf{B}\cdot\hat{n}|$, on the assumption that oppositely-directed flux tubes should not be able to reconnect and cancel.

From the flux we compute the magnetic critical mass
\begin{equation}
\label{eq:mphi}
m_{\Phi} = \frac{1}{2\pi} \left(\frac{\Phi}{\sqrt{G}}\right),
\end{equation}
and the magnetically-supported mass \citep{mouschovias76a, mckee07a}
\begin{equation}
m_B = \frac{m_\Phi^3}{m_{\rm gas}^2}.
\end{equation}
As with $m_{\rm BE}$ and $m_{\rm BE,10}$, we have this quantity as a function of radius for each sink particle.\footnote{We use $m_B$ rather than using $m_\Phi$ as our estimator of magnetic support  because $m_B$ is more analogous to $m_{\rm BE}$, in that both $m_B$ and $m_{\rm BE}$ are intensive quantities. The value of $m_{\rm BE}$ depends on the local density and temperature and thus does not change if we consider different volumes of constant density and temperature. Similarly. the ratio $m_B/m_{\rm gas}$ depends only on the mass to flux ratio $\Phi/m_{\rm gas}$, and thus does not vary if we consider volumes of varying size but fixed mass to flux ratio. In contrast, the total flux $\Phi$ and thus $m_\Phi$ are extensive quantities that do depend on the size of the volume considered, even if the conditions are uniform.} Note that the coefficient $1/2\pi$ used in \autoref{eq:mphi} depends on the exact density and magnetic field distribution, and can vary in the range $\approx 0.12 - 0.18$ \citep{tomisaka88a, tomisaka98a, mckee07a}. Our choice $1/2\pi \approx 0.16$ is appropriate for an infinite thin sheet \citep{nakano78a}, but lies near the upper end of the plausible range, and therefore likely gives an upper limit on the strength of magnetic support.

\autoref{fig:coreprof_example} shows an example of the types of profiles we generate, for a core around a 0.17 $M_\odot$ star in the strong magnetic field run. In addition to the density, we show the effective temperature and magnetic field, defined by
\begin{eqnarray}
\label{eq:teff}
T_{\mathrm{eff}} & = & \frac{\mu m_{\rm H}}{k_B} c_s^2 = \frac{\mu m_{\rm H}}{k_B}\frac{\int P\,dV}{m_{\rm gas}} \\
\label{eq:beff}
B_{\mathrm{eff}} & = & \frac{\Phi}{\pi r^2}. 
\end{eqnarray}
Thus $T_{\rm eff}$ and $B_{\rm eff}$ are the uniform temperature and magnetic field that would produce a sound speed and magnetic flux, respectively, equal to the mass-weighted mean sound speed and magnetic flux with radius $r$ that we measure. For $B_{\rm eff}$, we show the value derived using both our preferred definition of $\Phi$, where we take the absolute value of the field component normal to the surface and thus disregard cancellation of oppositely-directed flux tubes, and a result computing allowing cancellation, in order to illustrate that the difference between them is minor.

The final panel of \autoref{fig:coreprof_example} shows the profiles of the enclosed mass $m_{\rm gas}$, as well as the masses $m_{\rm BE}$, $m_{\rm BE,10}$ and $m_{B}$ that can be supported by thermal pressure with and without radiative heating, and by magnetic pressure, respectively.  From these measured profiles, we can compute three critical masses $m_{\rm BE,crit}$, $m_{\rm BE,10,crit}$ and $m_{B,\rm crit}$, defined as the masses enclosed within the shells where $m_{\rm BE}$, $m_{\rm BE,10}$ and $m_B$ are equal to $m_{\rm gas}$. That is, we find the intersection of the solid line in \autoref{fig:coreprof_example} with the three other lines. The physical meanings of these critical masses are clear. The enclosed mass $m_{\rm gas}$ is a strictly increasing function of radius and approaches 0 as $r\rightarrow 0$. The other masses remain finite as $r\rightarrow 0$, so that at sufficiently small $r$, $m_{\rm gas}$ is less than the masses that can be supported by thermal or magnetic pressure. Thus the gas at small $r$ is unable to collapse on its own and form another protostar; given its proximity to an existing star, it is likely to be accreted instead. On other hand, at sufficiently large radii the enclosed gas mass is greater than can be supported by thermal or magnetic pressure, and thus could at least potentially fragment to form another star rather than be accreted. Thus the critical masses provide rough estimates of the mass that will be added to the star by accretion. Large critical masses likely imply the formation of more massive stars, while small critical masses at least hold open the possibility of forming lower mass stars or brown dwarfs.

It is worth cautioning at this point that there are significant uncertainties in this calculation. The quantities we use to estimate the mass that can be supported against collapse -- $m_{\rm BE}$, $m_{\rm BE,10}$, and $m_B$ -- are computed for uniform gas without a central stellar source. A tidal gravitational potential due to the overall density gradient in the core and the presence of the central object will somewhat stabilise the gas. However, the effect is unlikely to be very significant. \citet{silk88a} analysed the stability of a Larson-Penston flow \citep{larson69a, penston69b} against linear radial perturbations and found that, once a central singularity (i.e., a star) forms, tidal effects do not prevent instabilities from growing linearly in time. Since this is case where tidal effects are likely at their strongest -- the perturbations they consider are purely radial and thus most easily suppressed by radial tides, and the tidal field for a Larson-Penston flow is maximally strong -- the true effect of tidal suppressing is likely to be even weaker.

Finally, note that we do not consider turbulent support in our analysis, a choice that is justified on observational, theoretical, and practical grounds. Observationally, the cores out of which low mass stars form are always observed to have significantly sub-thermal velocity dispersions, suggesting that turbulent support is unimportant \citep{goodman98b, pineda10a}. Theoretically, while turbulent support clearly can delay global collapse, it does not appear to be able to prevent collapse entirely, nor does it prevent fragmentation to small objects in those regions that do collapse. Thus for the purposes of determining a minimum mass scale for fragmentation, it is unclear that turbulent support should be included. Finally, as a practical matter the velocity fields in the vicinity of forming stars in our simulations are, not surprisingly, dominated by infall motions, and it is difficult to disentangle these from turbulence. Thus we cannot easily form a useful estimate of the amount of turbulent support from our simulations.

\subsection{Averaged Quantities}

Once we have computed the profiles and critical masses in the vicinity of every star, the next step is to compute mean values over all the stars in a given simulation. We bin the means by the mass of the star around which the core is found, so that we can study the evolution of profiles and critical masses with the mass of the central object. Formally, we measure a quantity $q$ at a series of output times $t_i$ in our simulations for each star present at that time; the stars have mass $m_{*,i,j}$, where $i$ indexes the output time and $j$ indexes the star at that time. We define the average of $q$ for a particular bin in central star mass $[m_{*,k}, m_{*,k+1})$ by
\begin{equation}
\label{eq:averages}
\langle q \rangle_k = 
\frac{\sum_i \sum_{j} q_{i,j} (t_{i+1}-t_{i})}{\sum_i \sum_{j} (t_{i+1}-t_{i})},
\end{equation}
where the sum over $j$ runs over all stars with mass $m_{*,i,j} \in [m_{*,k}, m_{*,k+1})$. Intuitively, $\langle q \rangle_k$ is an average over all the stars in a particular mass bin, with each star weighted by the time it spends in that mass bin. An alternative approach would be to average each object over the time it spends in a given mass bin and then average all objects equally. However, we prefer the approach of weighting in time because this most closely matches the average of what would be observed, since the probability of observing a particular state of evolution depends on its duration. Note that $q$ can be either a scalar quantity (e.g., $m_{\rm BE, crit}$) or a function of distance away from the star (e.g., $\rho(r)$). In all the analysis presented in this simulation, we use 15 logarithmically-spaced bins from $10^{-2} - 0.25$ $M_\odot$ for our analysis, thereby sampling from the mass where objects undergo second collapse to stellar density \citep{masunaga00a} up to the peak of the IMF. This is the phase we are interested in exploring, because it is during this phase that the peak of the IMF is determined.

\section{Results}
\label{sec:results}

\subsection{Mean Profiles of Density, Temperature, and Magnetic Field}

\begin{figure*}
\centerline{
\includegraphics[width=0.8\textwidth]{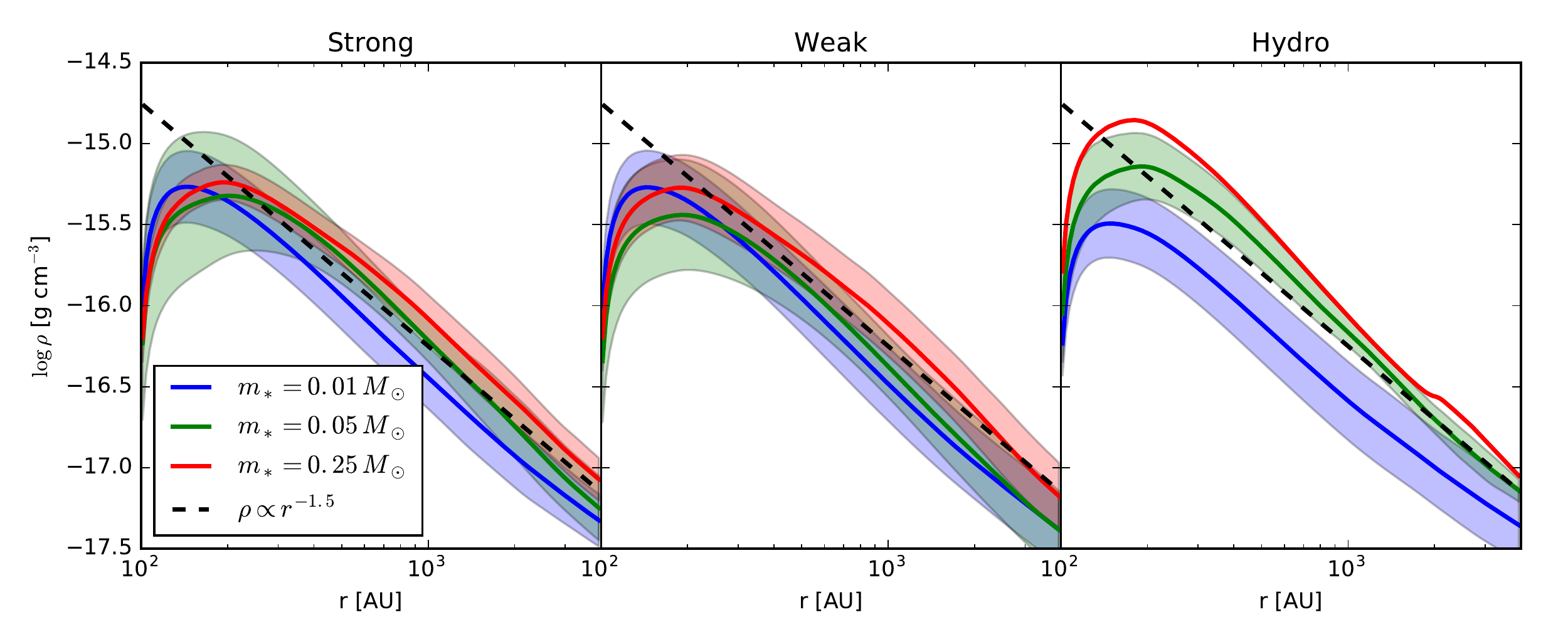}
}
\caption{
\label{fig:core_den}
Mean density profiles $\langle \rho(r)\rangle$ (defined per \autoref{eq:averages}) around protostars. Different panels show the results for the Strong magnetic field, Weak magnetic field, and Hydrodynamic runs, as indicated. Different colours show means for protostars of mass $0.01$ $\msun$, $0.05$ $\msun$, and $0.25$ $\msun$, respectively. For each mass bin, the central, thick line shows the mean, while the shaded band shows the $1\sigma$ dispersion of profiles in that mass bin. The shaded region is missing for the highest mass bin in the hydro run, because only a single star is in that bin, and thus a dispersion cannot be computed. The black dashed line, $\rho \propto r^{-1.5}$, is the same in every panel. It is not a fit, but is simply meant to guide the eye. The declines of the profiles inside $\sim 150$ AU are probably artificially imposed by the sink particle accretion and protostellar outflow injection algorithms, which alter the density profile on scales of a few computational zones; each zone is 23 AU in size.
}
\end{figure*}

\begin{figure*}
\centerline{
\includegraphics[width=0.8\textwidth]{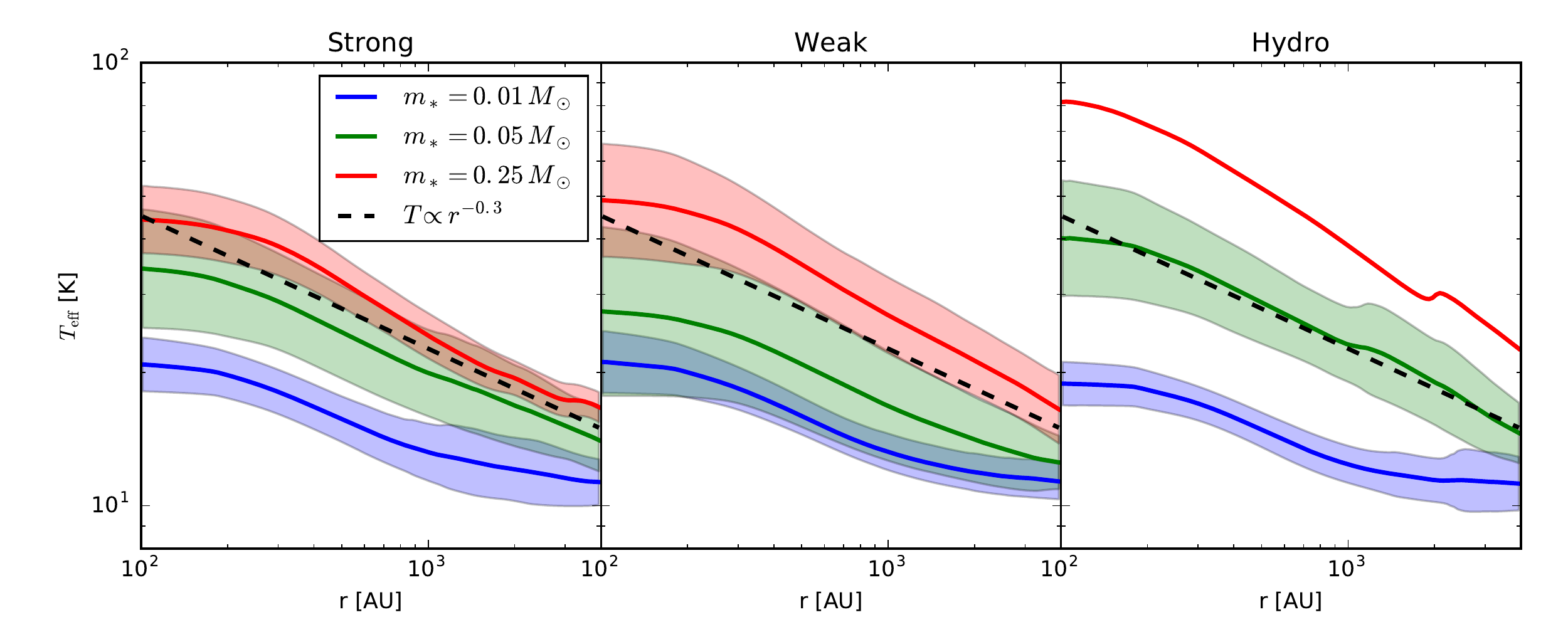}
}
\caption{
\label{fig:core_temp}
Same as \autoref{fig:core_den}, but showing the mean effective temperature $\langle T_{\rm eff}(r)\rangle$ (\autoref{eq:teff}) instead of the density. The black dashed line shows a scaling $T_{\rm eff} \propto r^{-0.3}$. This is not a formal fit, and is intended only to guide the eye. The initial temperature in the simulation volume is $T=10$ K.
}
\end{figure*}

\begin{figure*}
\centerline{
\includegraphics[width=0.6\textwidth]{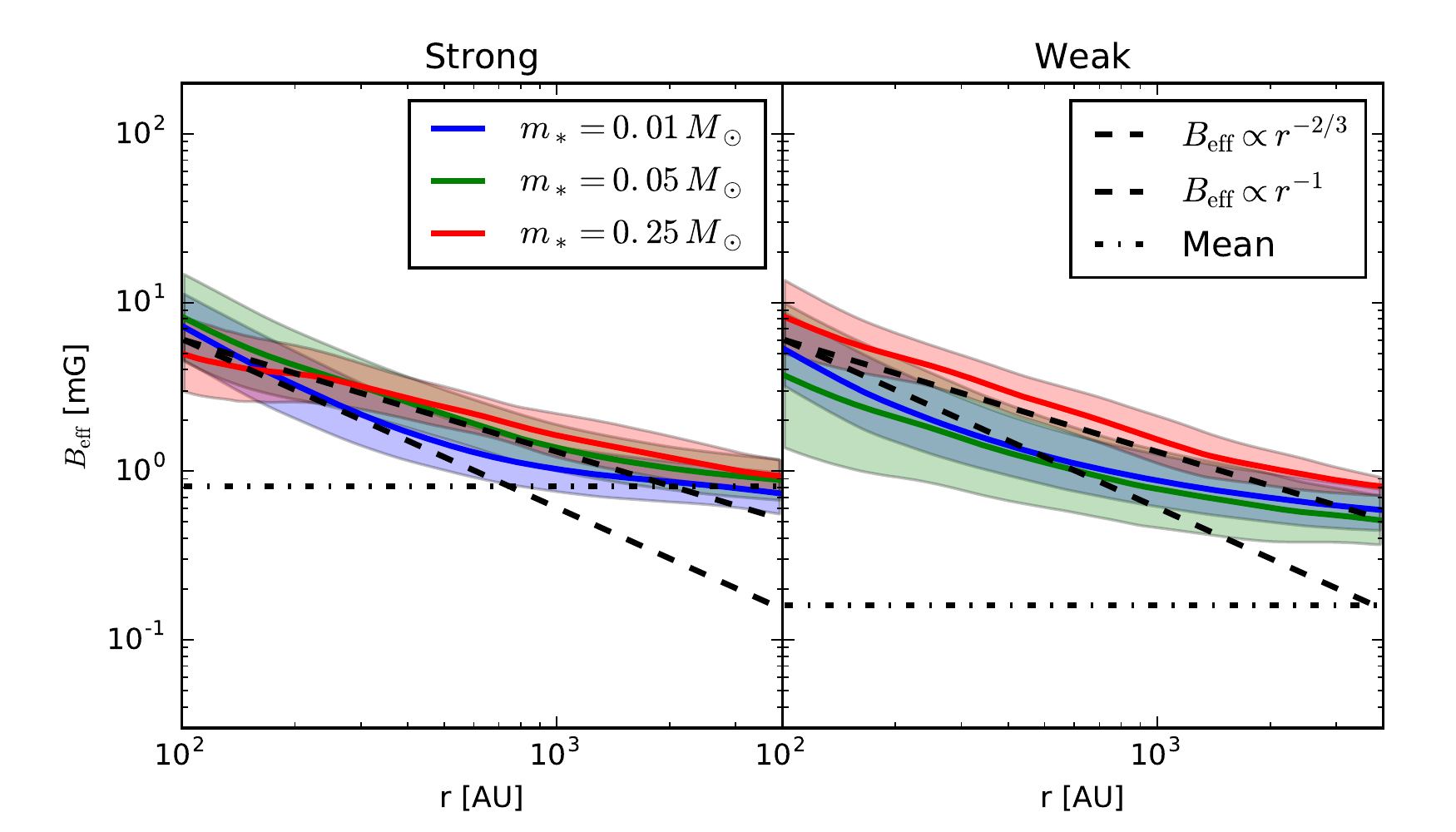}
}
\caption{
\label{fig:core_mag}
Same as \autoref{fig:core_den}, but showing the effective magnetic field strength $\langle B_{\rm eff}(r)\rangle$ (\autoref{eq:beff}). The Hydro run is omitted, since it has no magnetic fields. The dot-dashed horizontal lines show the effective field strength corresponding to the mean magnetic flux through the computational domain. The dashed lines, which are the same in both panels, show scalings $B_{\rm eff} \propto r^{-1}$ and $B_{\rm eff} \propto r^{-2/3}$. These are chosen to guide the eye, and are not formal fits.
}
\end{figure*}

We first examine the mean profiles of density $\langle \rho\rangle$, effective temperature $\langle T_{\rm eff}\rangle$, and effective magnetic field $\langle B_{\rm eff}\rangle$ in the vicinity of each protostar, as a function of protostellar mass. At this early phase of the evolution, when the protostar's mass is small, its final mass is likely to be determined by whether the gas around it accretes or fragments to form another star. We show the profiles we measure for the density, temperature, and magnetic field from all three simulations, and for three different central star masses, in \autoref{fig:core_den} (density), \autoref{fig:core_temp} (temperature), and \autoref{fig:core_mag} (magnetic field).

These figures enable a few immediate conclusions. First, examining \autoref{fig:core_den}, we find that cores have a density profile that is always close to $\rho^{-3/2}$, excluding the very smallest radii where the density drops due to numerical effects -- the accretion zone around each sink particle in our simulations is 92 AU in radius, and the protostellar injection region extends to roughly twice this size, so the density inside $\sim 150$ AU is artificially altered. In the Strong and Weak runs, the normalisation of the density profile is either non-evolving or very close to it as stars gain mass, while in the Hydro case it shows a weak increase with central object mass. The slope is consistent with what would be expected for free-fall collapse onto a point mass, but the non-evolution of the profile with mass is not, since for a Bondi-type flow the density at a fixed distance from the central object increases with mass. The density slope and the lack of evolution in the normalisation with central object mass (and thus, for a single object, with time) is consistent with the turbulent core model of \citet{mckee03a}, and with the model of \citet{murray15a} for the structure of a self-gravitating, turbulent, collapsing flow in the region near the centre of the collapse. Finally, we note that the densities are quite similar in all runs, indicating that the magnetic field has little effect on the density structure around cores. We shall see why this might be below.

\autoref{fig:core_temp} shows that the temperature, in contrast to the density, evolves significantly as protostars grow. This is not surprising: for a constant density profile, the accretion rate onto stars is increasing with time because infall velocities get larger. Since accretion luminosity is the dominant luminosity source for these low mass stars, they get brighter with time, and thus the temperature around them rises. However, note that there is non-trivial heating out to hundreds of AU even for the smallest stellar mass bin we consider, $\sim 0.01$ $M_\odot$. That heating is important even at such small masses was first pointed out by \citet{krumholz06b} based on analytic calculations, and by \citet{offner09a} and \citet{bate09a} using simulations. The slopes of the temperature profiles change with central object mass as well, being near $T_{\rm eff} \propto r^{-0.3}$ for the $0.05$ $\msun$ bin, somewhat shallower at lower central object masses, and somewhat steeper at higher masses. It is worth noting that these temperatures, while significantly elevated, are still nowhere near enough to present a barrier to accretion onto the central object. Even for $\sim 0.01$ $M_\odot$ central objects, the sound speed is a factor of several smaller than the escape speed from the gas and central star over the entire range of radii we consider, and this difference increases for the more massive central objects.

Finally, \autoref{fig:core_mag} shows the effective magnetic field strength (defined as the uniform magnetic field that would produce a flux equal to that we measure) in the two magnetised runs. It is interesting to note that the magnetic field profiles are quite similar in the two runs, despite the factor of 5 difference in the initial field strengths. In the strong field run, the effective magnetic field has fallen back to the background value by distances of $\sim 2000$ AU from the star, while in the weak field run it remains elevated over the background even out to 4000 AU, but the actual field strengths at small radii are similar. Both scale with distance as roughly $B_\mathrm{eff} \propto r^{-1}$ at small radii, flattening to closer to $B_{\mathrm{eff}}\propto r^{-2/3}$ at larger radii. The field does not increase significantly with protostellar mass. The lack of variation in the local field strength with either the large-scale magnetic field or the protostellar mass could have two possible causes. First, the field in the gravitationally-collapsing regions immediately around the protostars could be the result not of advection of field from large scales, but instead of a turbulent dynamo operating on small scales in the accretion flow around the stars \citep{sur12a, li15a}. This mechanism would amplify the field up to some saturation level, and would naturally explain why the field strength does not depend on the large-scale field. Second, the field could be dominated by flux advected into the region around the young stars, but this could be modulated by the escape of flux via magnetic interchange instability, whereby field lines that are drained of mass become buoyant and rise away from accreting protostars \citep{zhao11a, krasnopolsky12a, li14a}. This would explain the lack of growth of field with stellar mass. Regardless of the underlying mechanism, however, our result strongly suggests that the magnetic fields in the vicinity of young stars, those that are responsible for regulating fragmentation, are determined by a local process that is insensitive to the large-scale magnetic flux.

\subsection{Critical Masses}

\begin{figure*}
\includegraphics[width=0.9\textwidth]{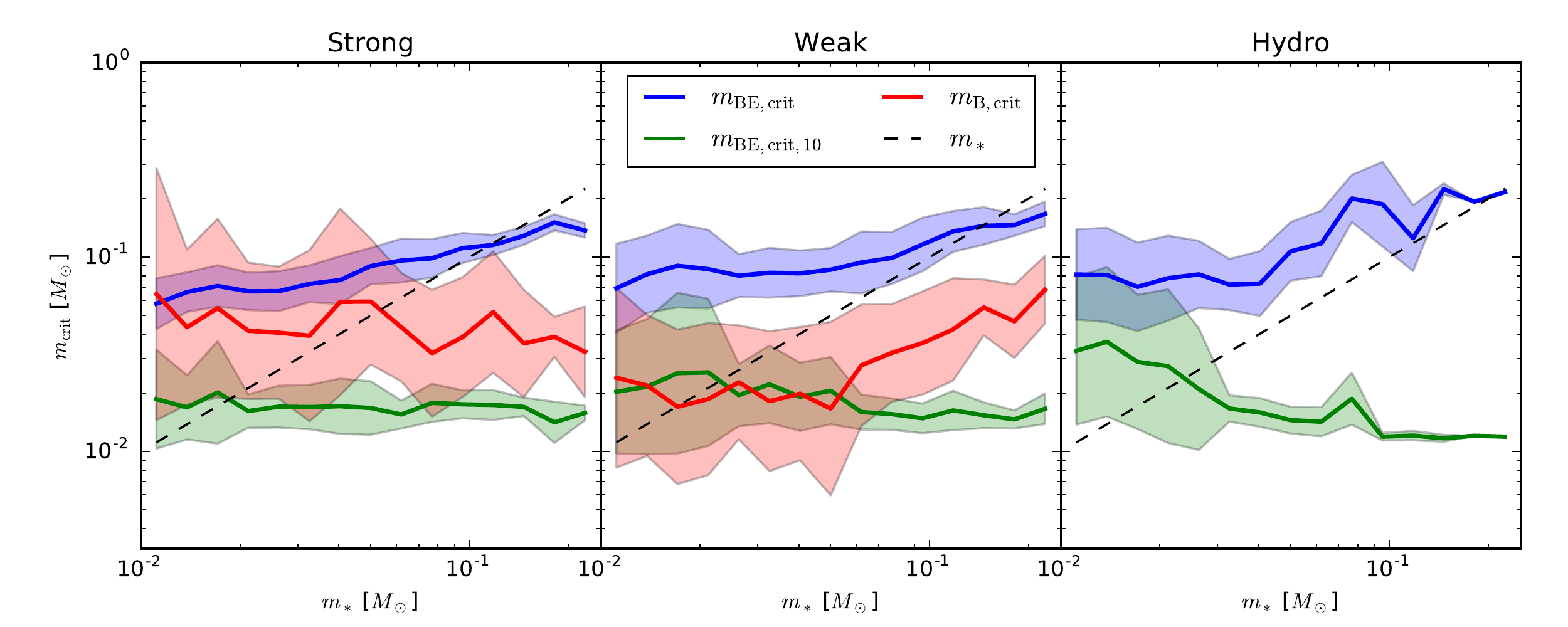}
\caption{
\label{fig:mcrit_core}
Critical masses of gas supported by thermal pressure ($m_{\rm BE}$, blue), thermal pressure at a fixed temperature of 10 K ($m_{\rm BE,10}$, green), and magnetic fields ($m_B$, red) as a function of central star mass $m_*$ in each of the three simulations, as labeled at the top of each panel. The shaded bands show the $1\sigma$ dispersion about each mean; places where the shaded band is absent indicate bins containing only a single star, so no dispersion can be computed. The dashed black line indicates $m_* = m_{\rm crit}$, i.e., it shows where the mass supported against collapse by pressure or magnetic forces is sufficient to double the present stellar mass.
}
\end{figure*}

We now ask how magnetic and thermal forces support gas against gravitational collapse. As discussed in \autoref{ssec:prof}, for each core at each time we can identify the radius and mass for which the enclosed mass is equal to the Bonnor-Ebert mass $m_{\mathrm{BE,crit}}$, the Bonnor-Ebert mass computed using a fixed temperature of 10 K $m_{\mathrm{BE,10,crit}}$, and the magnetic critical mass, $m_{B,\mathrm{crit}}$. We show the averages of these quantities as a function of central star mass in \autoref{fig:mcrit_core}.

From this plot we can immediately draw a few conclusions. First, comparing the lines for $m_{\rm BE}$ and $m_{\rm BE,10}$, we see that heating of the gas by stellar radiation dramatically increases the amount of mass that can be supported against fragmentation. This increase is a factor of $\sim 3$ even for the lowest mass stars near $0.01$ $M_\odot$, and becomes an order of magnitude by the time the star has grown to $\sim 0.2$ $M_\odot$. This dramatic suppression of fragmentation at early times by radiative heating is consistent with the arguments first made by \citet{krumholz06b}, and is not surprising given the suppression of fragmentation routinely seen in radiation-hydrodynamic simulations.

Perhaps more surprising is the relative unimportance of magnetic fields in providing support, even in the most strongly magnetised simulation we consider, and even with the relatively generous assumptions we have made in estimating magnetic support (e.g., ignoring cancelling magnetic fluxes, using a large coefficient in the computation of $m_\Phi$). While magnetic fields can support about half an order of magnitude more mass than could thermal pressure at a fixed temperature of 10 K, once we include radiative heating they provide an amount of support that is only comparable to thermal pressure for stars with mass $\lesssim 0.05$ $M_\odot$, and that is decidedly less important than thermal pressure for more massive stars. In the weak field run magnetic fields are less important than thermal pressure at essentially all stellar masses. The relative unimportance of magnetic support highlights the fact that the magnetic critical mass measured over large scales is not a particularly good guide to how important magnetic fields might be in shaping the IMF. In the strong field run, the magnetic field threading the entire computational domain is sufficient to prevent a mass $m_\Phi = 500$ $M_\odot$ from collapsing. However, the mass threading the few thousand AU-sized regions we are considering is far smaller than this, and can hold up far less mass. Turbulence in the simulations is able to gather mass along field lines and possibly also induce turbulent reconnection \citep{lazarian99a, santos-lima10a}, locally increasing the mass to flux ratio and creating regions where magnetic pressure becomes unimportant in comparison to thermal pressure.

Indeed, we note that the relative unimportance of magnetic fields as opposed to radiation in preventing fragmentation on small scales around stars, as opposed to the formation of additional stars far from existing ones, is consistent with the findings of radiation-magnetohydrodynamic simulations \citep{commercon11c, myers13a}. These show that radiation rather than magnetic fields is more important in prevent fragmentation close to growing stars. Magnetic fields are important in preventing the creation of new stars in low-density regions that are far from heating sources, but, once a region becomes unstable, they play little role in regulating the subsequent collapse and fragmentation.

Finally, we see that the thermal pressure including the effects of radiation feedback is the most effective mechanism for holding up the gas. It stabilises a gas mass in excess of the central star mass for all central stars smaller than $\sim 0.1$ $M_\odot$; the dispersion in $m_{\rm BE,crit}$ is surprisingly small. The typical $\sim 0.01$ $M_\odot$ star that has just formed due to a second collapse is immediately surrounded by an island of gas that is too hot to fragment, and which is $\sim 5$ times the mass of the star itself. As the star grows by accreting this gas, its luminosity rises and the heated island expands, remaining larger than the star until the star reaches $\sim 0.1$ $M_\odot$.

\section{Discussion}
\label{sec:discussion}

\subsection{Towards a Comprehensive Picture of the Origins of the IMF Peak}

\begin{figure*}
\includegraphics[width=0.9\textwidth]{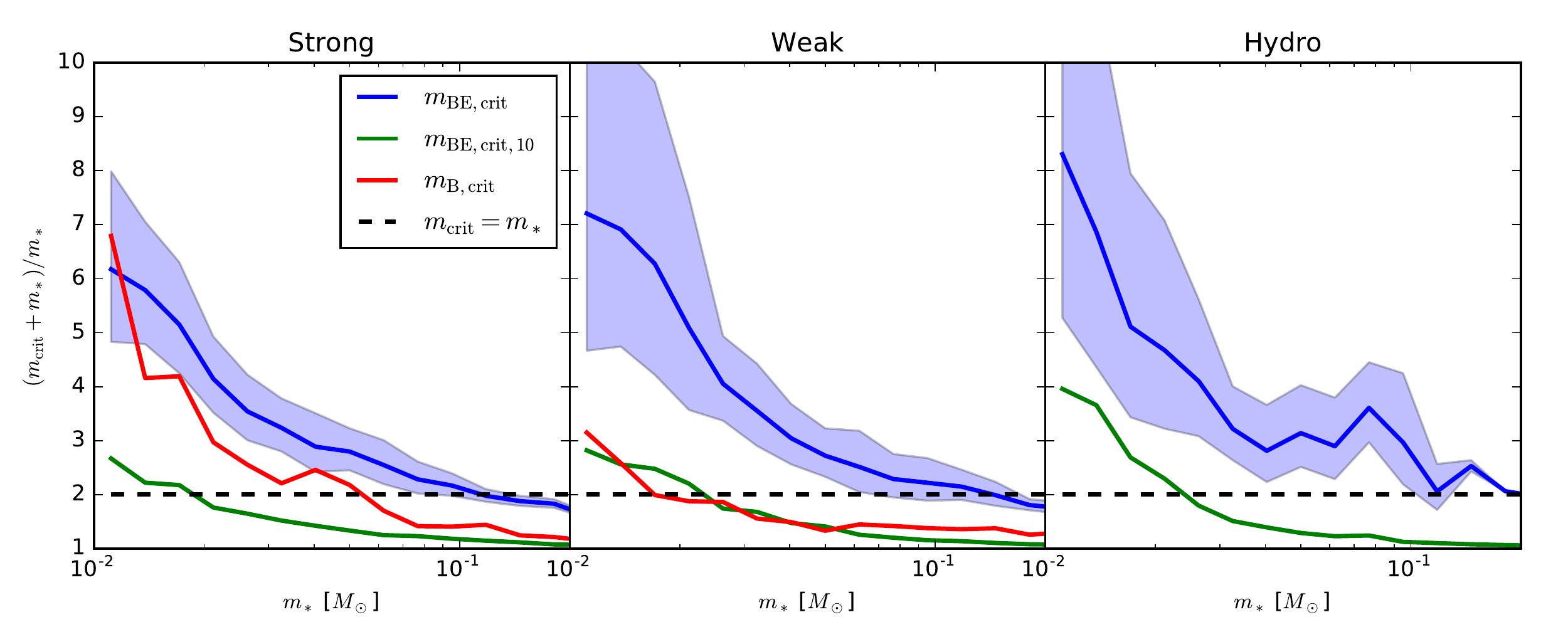}
\caption{
\label{fig:mcrit_ratio}
Ratio of stellar plus supported gas mass, $m_*+m_{\rm crit}$, to stellar mass alone, $m_*$, versus stellar mass. We show this quantity computed for the critical masses computed using the Bonnor-Ebert mass, $m_{\rm BE,crit}$ (blue), the Bonnor-Ebert mass using a gas temperature of 10 K, $m_{\rm BE,crit,10}$ (green), and the magnetically-supported mass, $m_{\rm B,crit}$ (red). For $m_{\rm BE,crit}$, solid lines show means, and shaded regions show the $1\sigma$ dispersion; we omit the shaded regions around $m_{\rm BE,crit}$ and $m_{\rm B,crit}$ to reduce clutter. The black dashed horizontal line shows $m_{\rm crit} = m_*$, i.e., it is the point at which the supported mass is sufficient to double the stellar mass.
}
\end{figure*}

The analysis we have performed is best understood within the context of a physical model for the origin of the IMF. Gas in star-forming clouds is able to cool on timescales much shorter than its dynamical time, and this makes it highly subject to fragmentation during collapse. Since the Jeans mass is a decreasing function of density as long as gas remains isothermal or close to it, this fragmentation proceeds to ever-smaller mass scales. If this process continued unimpeded, the resulting mass function would peak near the opacity limit for fragmentation, $\sim 0.004$ $M_\odot$ \citep{low76a, rees76a, whitworth07a}. The actual peak of the IMF, which is $\sim 2$ orders of magnitude larger than this, is determined by whatever arrests this cascade of fragmentation. Put more succinctly, it is helpful to rephrase the question ``what sets the peak of the IMF?" as the question ``what suppresses the formation of brown dwarfs?"

Since the equations of isothermal self-gravitating magnetohydrodynamics do not by themselves impose a mass scale \citep{mckee10b, krumholz14c}, suppression of the formation of low mass objects must come from either the initial conditions or from a deviation from isothermality. An example of the former approach is the model of \citet{padoan04a}, who posit that brown dwarfs are rare because only in the most unusual, high-pressure regions is it possible for a region of gas in the brown dwarf mass range to become gravitationally bound. Since the requirements for becoming bound are ultimately set by the strength of the turbulence, as parameterised by the normalisation of the linewidth-size relation, in this class of models the frequency of brown dwarfs should depend on this normalisation. This is potentially problematic, since one then predicts a significant overabundance of brown dwarfs relative to stars in regions with stronger turbulence at fixed size scale, as is typically found in massive star-forming clumps \citep{shirley03a}. At present there is no evidence for such variation in the brown dwarf to star ratio, but further observations are needed to rule it out.

Our findings suggest an alternate explanation for the paucity of brown dwarfs relative to stars. We find that fragmentation in the immediate vicinity of young stars is suppressed primarily by thermal radiation feedback from the star itself. As noted above, stars with masses below the peak of the IMF, $\sim 0.1$ $M_\odot$ or less, are invariably surrounded by a mass of gas less than a thermal Jeans mass that is in excess of the mass of the central star. This would not be true in the absence of stellar radiation. In simulations including a strong initial magnetic field, magnetic support is comparable to thermal support in importance for objects up to $\sim 0.05$ $M_\odot$, but is substantially less important by the time the central object reaches $\sim 0.1$ $M_\odot$, suggesting that thermal support is ultimately the more important process.

To illustrate this, in \autoref{fig:mcrit_ratio} we show the ratio $(m_{\rm BE,crit}+m_*)/m_*$ (blue line), i.e., this is the fraction by which the mass of the star would be increased by accretion of all the material around itself that is too warm to fragment. This is $\gtrsim 5$ at the lowest masses, which naturally explains why brown dwarfs are comparatively rare: a ``prospective" dwarf of mass $\sim 0.01$ $M_\odot$ is usually luminous enough to have heated $\sim 0.05$ $M_\odot$ of material around itself to a point where it is too hot to collapse, and instead seems very likely to be accreted. By the time the object has accreted this gas and grown to $\sim 0.05$ $M_\odot$, it has heated up another $\sim 0.05$ $M_\odot$ of material to the point where it cannot fragment, enabling it to grow to $\sim 0.1$ $M_\odot$, and so forth. This process continues but comes ever less important as stars gain in mass. By the time stars approach the peak of the IMF, $\sim 0.2-0.3$ $M_\odot$, the amount of heated mass around them has fallen to tens of percent of their current mass, and represents a relatively minor perturbation if and when it is accreted, particularly since protostellar outflows are likely to eject $\sim 50\%$ of it \citep{matzner00a}. Conversely, because stars much above $\sim 0.2-0.3$ $M_\odot$ are unable to stabilise enough mass around themselves to significantly augment their mass, they seem likely to be starved by fragmentation of this gas into other stars, as suggested by \citet{peters10b}. This explains why we should expect the peak of the IMF to fall at $\sim 0.2-0.3$ $M_\odot$. Only in rare circumstances does this heating mechanism allow an object to remain at $\sim 0.01$ $M_\odot$, rather than continuing to grow. It is worth noting here the analogy between this explanation for the rarity of brown dwarfs and the analysis of giant planet formation by disc instability by \citet{kratter10b}, who show that disc instability can happen, but that the objects it creates usually wind up as binary companions rather than ceasing accretion at planetary masses.

We pause to note that the fact that nothing in our calculation pre-ordained that $(m_{\rm BE,crit} + m_*)/m_*$ had to become of order unity at a mass of $\sim 0.2$ $M_\odot$. There is no obvious reason why we could not have found that this transition occurs at, for example, $\sim 0.01$ $M_\odot$. Indeed, examining \autoref{fig:mcrit_core}, this is precisely what we do find if we omit radiation. The fact that our calculation including radiation moves this special mass to exactly where the peak of the IMF is observed to lie is highly suggestive of the importance of radiative feedback.

\subsection{Accretion Stopping}
\label{ssec:accretion_stop}

Our discussion to this point has assumed that the material within a few hundred AU of a protostar that is unable to fragment will actually accrete onto the star and be able to raise its mass. Is this a fair assumption? The stabilised regions that we identify are quite small, typically $\sim 500$ AU in radius, and have velocities that are predominantly inward toward the central star. At the typical densities of $\sim 10^{-16}$ g cm$^{-3}$ in these regions (\autoref{fig:core_den}), the time required for free-fall collapse is $\lesssim 10$ kyr; if the stellar mass is of order the gas mass, then the collapse time will be further reduced. Thus any mechanism that could potentially interfere with accretion must be capable of acting on such a timescale. We consider here two possibilities.

\subsubsection{Dynamical Interactions}

One way that accretion could be stopped is if two accreting stars get close enough to one another for one of the stars to be ejected from its stabilised gaseous core, or for one star to capture a significant amount of mass from the other's stabilised region. To check whether this is likely to happen, we note that the typical stellar density in these simulations is $n_*\lesssim 10^4$ pc$^{-3}$ (see Figure 16 of \citealt{myers14a}). For stars moving at the velocity dispersion of $v = 1.2$ km s$^{-1}$ used in the simulations, the mean time required for two of these 500 AU regions to encounter one another (using a cross section $\sigma = \pi (1000\mbox{ AU})^2$ is $t = 1/(n_*\sigma v) \sim 1$ Myr.

Thus the typical star in our simulations will not experience an encounter that is likely to inhibit its ability to accrete the stabilised gas around it. There are exceptions; the upper envelope of stellar density in the simulations extends to $\sim 10^6$ pc$^{-3}$, and in these cases the encounter time is within a factor of a few of the collapse time. Indeed, we have argued that these rare cases where stars are very close to one another are likely the source of brown dwarfs. Nonetheless, this analysis reinforces our conclusion that dynamical interactions are unimportant for the typical star. We note that this conclusion is also consistent with observations of the kinematics of protostellar cores \citep[e.g.,][]{andre07a}, which also generally find that their free-fall times are short compared to the timescales that would be required for them to interact dynamically.

\subsubsection{Photoevaporation and Photodispersal}
\label{ssec:photoevap}

One other way that stars could be prevented from accreting their stabilised regions is if they are removed by radiative processes. The region simulated in \citet{myers14a} is modelled after the Orion Nebular Cluster, and has enough mass and density that, if the simulations had been continued long enough, it should eventually have produced massive stars. These stars in turn would produce ionising radiation. This will halt formation of additional stars, which poses no problem for our model, but if it is also able to remove the stabilised material around existing stars and prevent it from accreting, that does present a problem because this would leave the region with a bottom-heavy mass function comparable to the protostellar mass function, rather than something similar to the IMF.

Consider placing one of our stabilised regions at a distance $R_*$ from a massive star with an ionising photon luminosity $Q$. When the ionising photons first hit the stabilised region, they will immediately photoionise its outer surface. The depth $L$ of this photoionised layer can be estimated by balancing the recombination and ionisation rates per unit area:
\begin{equation}
\alpha^{\rm (B)} n_e n_{\rm H} L \approx \frac{Q}{4\pi r_*^2},
\end{equation}
where $\alpha^{\rm (B)} \approx 3\times 10^{-13}$ cm$^3$ s$^{-1}$ is the case B recombination coefficient. Inverting this, and plugging in distance of $R_* = 1$ pc (roughly the half-mass radius of the ONC -- \citealt{hillenbrand98a}), an ionising luminosity $Q = 10^{49}$ s$^{-1}$ (roughly the ionizing luminosity of $\theta^1$ Ori C -- \citealt{draine11a}), and a characteristic density $n_e \approx n_{\rm H} \approx 10^8$ cm$^{-3}$ for the thermally-stabilised material around each protostar at a distance of $\approx 500$ AU (see \autoref{fig:core_den}), we have
\begin{equation}
L \approx \frac{Q}{4\pi \alpha^{\rm (B)} n_e n_{\rm H} r_*^2} \approx 2\times 10^{-6} Q_{49} n_8^{-2} R_{*,0}^{-2}\mbox{ AU},
\end{equation}
where $Q_{49} = Q/10^{49}$ s$^{-1}$, $n_8 = n/10^8$ cm$^{-3}$, and $R_{*,0} = R_*/1$ pc. Evidently, the photoionized layer is very thin compared to the size of the region in question.

The thin photoionised layer will be heated to a temperature of $\approx 10^4$ K. Since this is far greater than the escape speed from the star or its core, this material will rocket off in a freely-expanding wind, and the stablised region will begin to ablate. The back-reaction from this flow will both compress and accelerate the cloud. If the ionising radiation is able to evaporate the stabilised region or drive it away from the star on a timescale short compared to the time required for it to be accreted, the photoionisation can halt accretion.

\citet{bertoldi89a} and \citet{bertoldi90a} solve the problem of photoevaporation and rocket acceleration for uniform clouds without gravity, and we therefore use their solution to set an upper limit on the effects of these phenomena for our gravitationally-confined regions. \citeauthor{bertoldi90a} show that a cloud of mass $m$, initial density $n$, and initial magnetic field strength $B$ that is exposed to a planar ionising flux will develop an equilibrium cometary structure with a steady mass flow off its surface. The approximate radius of this structure is, from \citeauthor{bertoldi90a}'s equation (3.31),\footnote{For simplicity we have set \citeauthor{bertoldi90a}'s dimensionless parameters $\phi_{\rm mr}$, $\phi_m$, $\phi_D$, and $\phi_{4/3}$ to exactly unity in what follows; since we are interested in an order of magnitude estimate, this is sufficient. We set $\omega=0.13$, from \citet{bertoldi89a}'s Table 1, Models 13 and 14.}
\begin{equation}
\label{eq:rphoto}
r = 970 \left(\frac{m_0 B_0^{3/2} R_{*,0}^{3/4}}{n_8 Q_{49}^{3/8}}\right)^{8/21}\mbox{ AU},
\end{equation}
where $m_0 = m/M_\odot$ and $B_0 = B / 1$ mG. Plugging in the values $m_0 = 0.1$, $n_8 \approx 1$, and $B_0 = 2$ that characterise our stabilised regions at $\sim 500$ AU (\autoref{fig:core_mag}), we have an equilibrium radius of $\approx 600$ AU, nearly identical to this size of the stabilised region even absent photoionisation. In this configuration, the equilibrium mass loss rate is (\citeauthor{bertoldi90a}'s equation (4.3b))
\begin{equation}
\dot{m} = 4.8\times 10^{-7} m_0^{4/7} \left(\frac{B_0}{n_8^{2/3}}\right)^{6/7} \left(\frac{Q_{49}}{R_{*,0}^2}\right)^{2/7}\,M_\odot\mbox{ yr}^{-1},
\end{equation}
giving a value $2.3\times 10^{-7}$ $M_\odot$ yr$^{-1}$ for our fiducial parameters. Thus the time required to photoevaporate the stabilised region is $\approx 430$ kyr, more than an order of magnitude longer than the $\lesssim 10$ kyr required for the region to collapse. Thus we see that photoevaporation is unlikely to be able to strip away a significant amount of the mass.

In addition to evaporating the mass, the ionisation could also disperse it. The evaporating gas will rocket away at $c_i \sim 10$ km s$^{-1}$, and this will exert a force back on the neutral gas, producing a D type ionisation front that will both implode the stabilised region and accelerate it away from the ionising source. Only the latter process could interfere with the gas being accreted; implosion will, if anything, cause accretion to occur more rapidly. From \citeauthor{bertoldi90a}'s formalism, the velocity to which the cloud is accelerated is
\begin{equation}
v_c \approx 0.0077 c_i m_0^{-1/12} n_8^{-5/12} \left(\frac{Q_{49}}{R_{*,0}^2}\right)^{1/4},
\end{equation}
which gives $v_c\approx 0.08$ km s$^{-1}$ for our fiducial parameters. In comparison, the escape speed at a distance of 500 AU from even a $0.01$ $M_\odot$ central object is $0.19$ km s$^{-1}$. Thus ionising radiation does not accelerate the neutral material to a speed that is high enough to escape and avoid being accreted. Even if we ignore this effect, the time required for the gas to move away from the central star at this speed would be $t_{\rm disp} \approx 500\mbox{ AU}/v_c \approx 30$ kyr, which is again significantly longer than the time required for the gas to be accreted, even ignoring acceleration of the accretion due to implosion. Thus we conclude that the presence of a star with a significant ionising flux cannot plausibly prevent accretion of the stabilised regions either by photoevaporating or dispersing them, and therefore that our assumption that this material is bound to accrete if it does not fragment is reasonable.

\subsection{Caveats and Cautions}

We end this discussion by cautioning that the simulations on which we base our conclusions use initial conditions chosen to be appropriate for star formation in the Milky Way. It is an open question whether the same results would hold for radically different conditions, for example those found in a low-metallicity dwarf galaxy or in a starburst or high redshift galaxy where the gas is much denser and more turbulent than is typical of the Milky Way. Radiation-hydrodynamic simulations of gas fragmentation and star formation at varying metallicity suggest that our results should be robust against metallicity variations at least over several dex in metal abundance \citep{myers11a, bate14a}, but fully answering that question will require repeating the analysis presented here at a range of metallicities.

On the question of how gas density and turbulence might affect the results, there are no simulations in the literature that offer much guidance, only analytic and semi-analytic models. The density of the gas will affect the typical accretion rate onto stars, which in turn will change their luminosities, enabling them to heat more material. On the other hand, increasing density makes it easier for gas to fragment by lowering the Bonnor-Ebert mass. \citet{krumholz11e} and \citet{guszejnov16a} find that these two effects nearly cancel, yielding a characteristic mass that is close to independent of density. However, this proposition remains untested by simulations.

Finally, we emphasise that the findings we present here pertain to the location of the peak of the IMF. They do not address the origin of the powerlaw tail that extends to high masses. This feature of the IMF has been ascribed by various authors either to a second phase of competitive accretion that follows the initial formation of small seeds \citep[e.g.,][]{zinnecker82a, bonnell97a, bonnell01a, bonnell01b}, or to the rare turbulent generation of massive structures that are bound, but that have little enough substructure within them that they fragment minimally or not at all as they collapse \citep[e.g.,][]{padoan02a, mckee03a, hennebelle08b, hennebelle09a, hopkins12d, guszejnov15a}. Because we do not form stars with masses significantly above the peak of the IMF in our simulations, we are unable to address which, if either, of these models correctly accounts for the IMF's powerlaw tail.

\section{Summary and Conclusion}
\label{sec:conclusion}

We have investigated the structure of the gas in collapsing regions formed in simulations of the formation of star clusters. Because the simulations include a very wide range of physics -- turbulence, magnetohydrodynamics, gravity, radiative transfer (including radiation from young stars), and protostellar outflows -- we are able to isolate the collapsing regions and compare the influence of various processes in determining where gas does and does not fragment, and thus in setting the peak of the IMF.

We find that collapsing cores approach a profile of density versus distance from the central object that is constant in time, with no dependence on the initial magnetic field strength in the simulations. The same is true of the magnetic flux, strongly suggesting that the magnetic flux in the vicinity of a collapsing core is determined more by local processes than by the flux present on larger scales. The temperature of the gas, on the other hand, strongly varies with central star mass, rising with time as stars grow and accrete more rapidly. The relatively small amount of magnetic flux around young stars, coupled with the rising temperatures, means that the gas in the immediate vicinity of young stars is primarily prevented from collapsing on its own (as opposed to accreting onto the already-collapsed object) by thermal pressure. 

This support is non-negligible for protostellar objects in the brown dwarf mass range. The amount of mass around the object that is too warm to be able to fragment is several times the mass of the object itself, and this gas will likely be accreted unless some external event intervenes. This will push the object into the stellar mass range, a phenomenon that explains why brown dwarfs are rare compared to stars. This effect ceases to be significant once stars reach $\sim 0.2$ $M_\odot$, because stars at this mass can only suppress fragmentation in a mass that is only a few tens of percent of their own, not enough to make the star grow significantly. This finding strongly suggests that radiation feedback is the key process in determining the location of the peak of the IMF. Radiative heating suppresses formation of stars below the peak, and then stops operating, leaving the majority of stars with masses $\sim 0.2$ $M_\odot$ as we observe.

\section*{Acknolwedgements}

MRK acknowledges support from Australian Research Council grant DP160100695. MRK, RIK, and CFM acknowledge support from NASA TCAN grant NNX-14AB52G and NASA ATP grant NNX-13AB84G. CFM and RIK acknowledge support from NSF grant AST-1211729. RIK acknowledges support from the US Department of Energy at the Lawrence Livermore National Laboratory under contract DE-AC52-07NA27344. The simulations reported in this paper made use of the pleiades supercomputer at NASA Ames, through a grant of time awarded under NASA ATP grant NNX-13AB84G.

\bibliographystyle{mn2e}
\bibliography{refs}

\begin{appendix}

\section{Cumulative versus Differential Quantities}
\label{app:differential}

\begin{figure}
\includegraphics[width=\columnwidth]{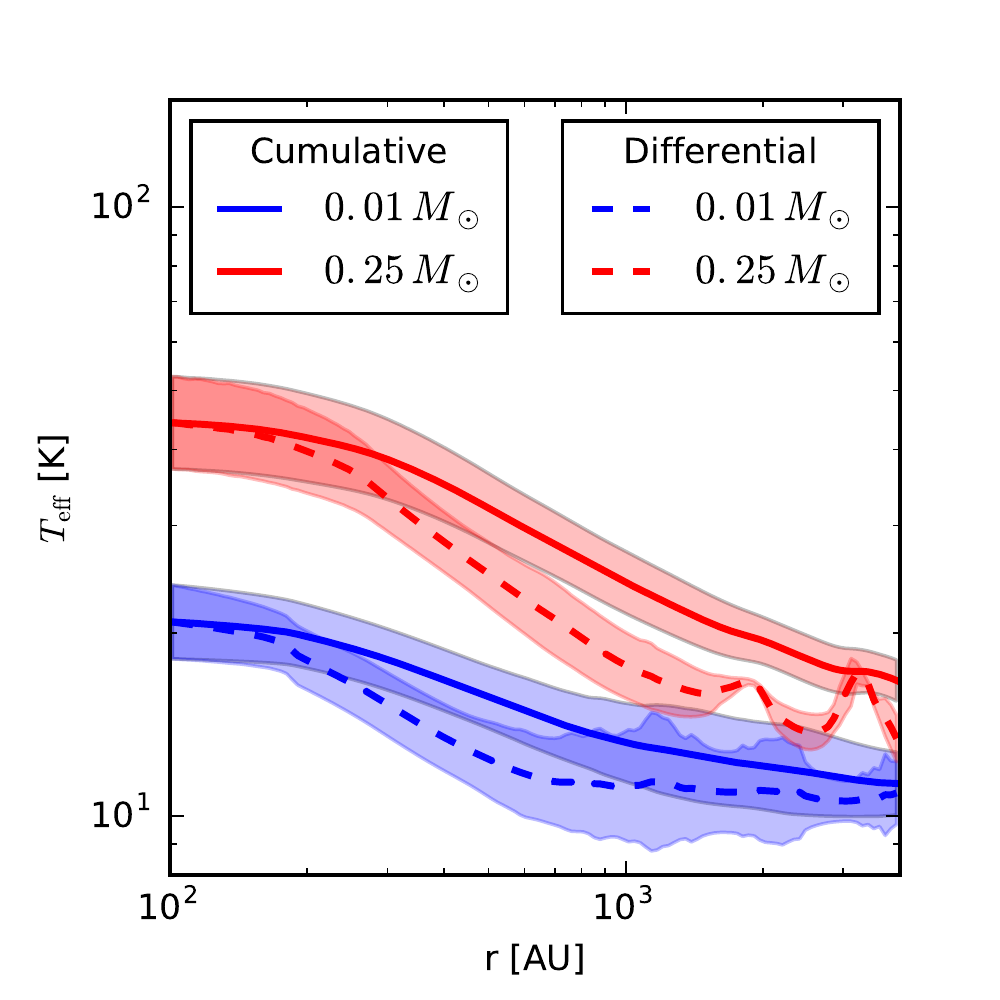}
\caption{
\label{fig:temp_diff}
Mean effective temperature $\langle T_{\rm eff}\rangle $ versus distance from the central star computed  both cumulatively (i.e., using all the mass inside a given radius -- solid line) and differentially (i.e., just using the mass in a thin shell at a given radius -- dashed line). The values shown are for the strong magnetic field run, and averages in two different mass bins, centred on $0.01$ $M_\odot$ (blue) and $0.25$ $M_\odot$ (red). Shaded regions show the $1\sigma$ scatter in the means.
}
\end{figure}

\begin{figure}
\includegraphics[width=\columnwidth]{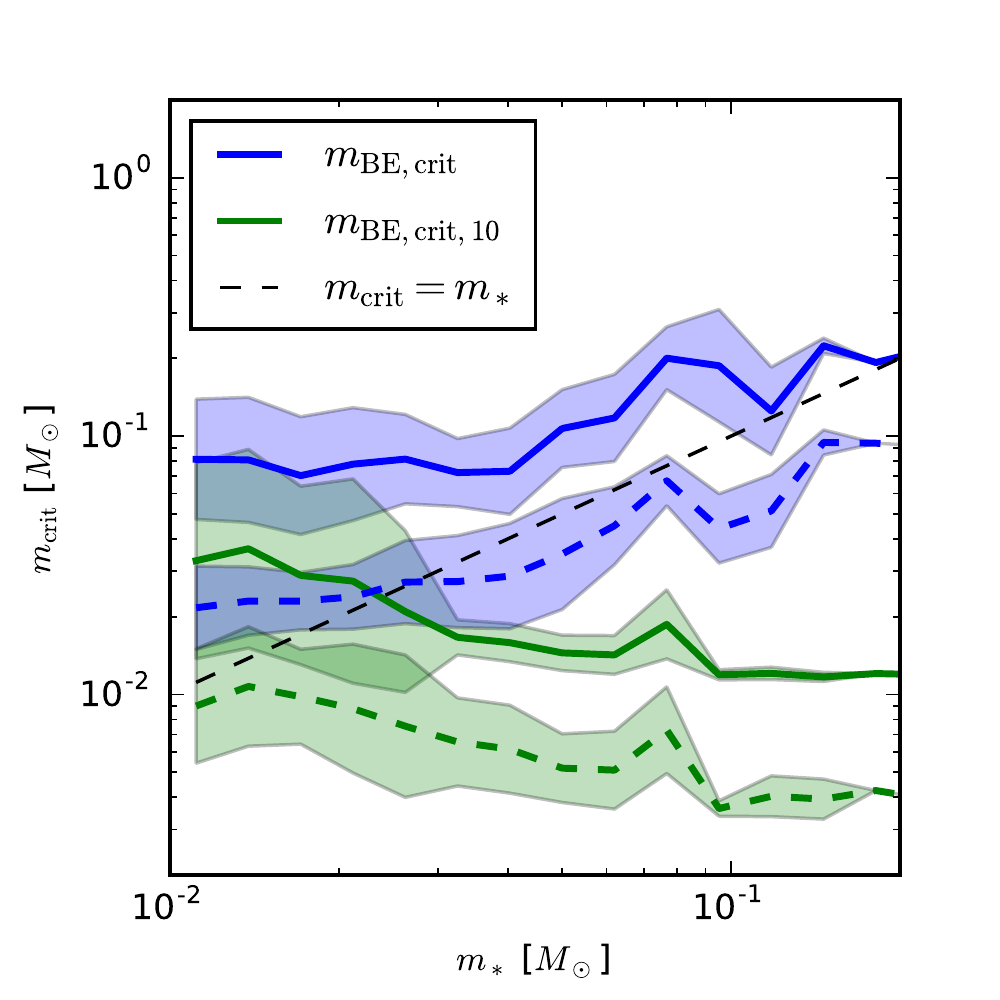}
\caption{
\label{fig:mcrit_diff}
Mean critical mass supported by thermal pressure $\langle m_{\rm BE,crit}\rangle$ (blue) and by thermal pressure at a fixed temperature of 10 K $\langle m_{\rm BE,10,crit}\rangle$ (green), versus central star mass $m_*$. We show quantities computed both cumulatively (i.e., using all the mass inside a given radius -- solid line) and differentially (i.e., just using the mass in a thin shell at a given radius -- dashed line). The values shown are for the strong magnetic field run. Shaded regions show the $1\sigma$ scatter in the means. We omit plots of $\langle m_\phi\rangle$ to avoid cluttering the plot.
}
\end{figure}

As discussed in \autoref{ssec:prof}, the mass-weighted mean sound speed and other quantities that enter calculation of the critical masses can be computed cumulatively, meaning taking the mass-weighted mean of all material inside a given radial shell, or differentially, meaning computed using only the material between two adjacent shells. The cumulative choice seems more physically reasonable for the purposes of computing the mass supported against collapse, since the calculation of the mass enclosed within a given radius is necessarily cumulative. Nonetheless, for completeness we have repeated all of the analysis included in the main text using the differential definition. \autoref{fig:temp_diff} shows the effective temperature in the strong magnetic field run using the differential definition, and \autoref{fig:mcrit_diff} shows the critical masses; note that for the differential case, we use a coefficient of $1.18$ rather than $1.86$ in \autoref{eq:mbe}. The other runs show similar results, and are omitted for reasons of space.

As the plots show, the differential effective temperature is, not surprisingly lower. The differential density (not shown) is somewhat higher, due to omission of the central evacuated region. The net result is that both the heated and non-heated critical masses are somewhat reduced compared to the cumulative definition. However, the result that the critical mass with heating is higher than that without heating and than the magnetically-supported mass, and that it exceeds the mass of the central object in the brown dwarf mass regime, continues to hold.

\end{appendix}

\bsp

\label{lastpage}

\end{document}